\documentclass{PoS}

\usepackage[numbers,compress]{natbib}
\usepackage{xcolor}
\usepackage{graphicx}

\usepackage[hang,flushmargin]{footmisc}

\usepackage{amsmath}
\usepackage{cancel}
\usepackage{slashed}

\usepackage{ucs}
\usepackage[T1]{fontenc}
\usepackage{lmodern}

\colorlet{yellow}{yellow!70!black}

\newcommand\Dslash{\slashed D}
\newcommand\pslash{\slashed p}
\newcommand\Tr{\mathop{\rm Tr}}
\newcommand\diag{\mathop{\rm diag}}

\makeatletter
\def\Hy@colorlink#1{\begingroup}
\def\Hy@endcolorlink{\endgroup}
\usepackage[unicode,hyperfootnotes=false]{hyperref}
\newcount\Hy@sublinkcounter
\newcommand\phantomsubsection{%
  \Hy@GlobalStepCount\Hy@sublinkcounter
  \xdef\@currentHref{subsection*.\the\Hy@linkcounter.\the\Hy@sublinkcounter}%
  \Hy@raisedlink{\hyper@anchorstart{\@currentHref}\hyper@anchorend}}
\ifx\phantomsection\undefined\newcommand\phantomsection{}\fi
\expandafter\renewcommand\expandafter\phantomsection\expandafter
  {\phantomsection\global\Hy@sublinkcounter0\relax}
\@ifundefined{texorpdfstring}
{}{}

\@ifundefined{unichar}
{\newcommand\myunichardef[3]{\expandafter\providecommand\csname text#1\endcsname
                            {#2}}}
{\newcommand\myunichardef[3]{\expandafter\providecommand\csname text#1\endcsname
                            {\unichar{"#3}}}}
\makeatother
\myunichardef{composetilde}{\string~}{0303}
\myunichardef{composemacron}{bar}{0305}
\myunichardef{subscriptthree}{\string_3}{2083}

\newcommand\footpunct[2]{{\hbox to 0pt\expandafter{\expandafter\spacefactor\the\spacefactor\relax%
    #2\relax\expandafter\hss\expandafter}\expandafter\def\expandafter\tmp\expandafter{\expandafter\spacefactor\the\spacefactor\relax}%
    \footnote{#1}\tmp}}

\newbox\dpbox
\setbox\dpbox\hbox{}
\dp\dpbox=3\jot

\title{Neutron Electric Dipole Moment from Beyond the Standard Model}

\ShortTitle{nEDM from BSM}

\author{\speaker{Tanmoy Bhattacharya}\\
        Los Alamos National Laboratory\\
        E-mail: \email{tanmoy@lanl.gov}}

\author{Boram Yoon\\
        Los Alamos National Laboratory\\
        E-mail: \email{boram@lanl.gov}}

\author{Rajan Gupta\\
        Los Alamos National Laboratory\\
        E-mail: \email{rajan@lanl.gov}}

\author{Vincenzo Cirigliano\\
        Los Alamos National Laboratory\\
        E-mail: \email{cirigliano@lanl.gov}}

\abstract{We present an update on our calculations of the matrix
  elements of the CP violating quark and gluon chromo-EDM operators,
  as well as the operators these mix with, such as the QCD
  Theta-term. Their contribution to the neutron EDM is obtained by
  extrapolating the \(F_3\) form factor of a vector current to zero
  momentum transfer. The calculation is being done using valence
  Wilson-clover quarks on HISQ background configurations generated by
  the MILC collaboration.}

\FullConference{The 36th Annual International Symposium on Lattice Field Theory - LATTICE2018\\
		22-28 July, 2018\\
		Michigan State University, East Lansing, Michigan, USA.}

\begin{document}

\section{Introduction}
\vspace*{-0.5\baselineskip}

Violation of the symmetry under simultaneous charge-conjugation and
parity-flip (CP) is a core ingredient in the standard model (SM) and
is necessary to explain the vast excess of matter over antimatter in
the universe~\citep{0038-5670-34-5-A08}.  The SM CP-violation (CPV) is
small and arises from the weak mixing between the
quark~\citep{1963PhRvL..10..531C,1973PThPh..49..652K}, and possibly
also lepton~\citep{Pontecorvo:1957qd,1962PThPh..28..870M},
families. Cosmological models require much stronger
CPV~\citep{Trodden:1998ym}, and most theories beyond the SM (BSM) do
indeed produce it naturally.  If this additional CPV is produced
naturally by physics at a few TeV, the next generation of electric
dipole moment (EDM)
measurements~\citep{2011JPhCS.335a2012S,Chupp:2017rkp} are likely to
find it, and the neutron is a very good candidate system.  To connect theory to 
experiments, it is imperative to obtain the matrix elements (ME) of the CPV
effective operators that control the EDMs of various particles.  Here,
we discuss our progress in calculating the neutron EDM (nEDM) due to
the quark chromo-EDM operators.\looseness-1

\subsection{BSM Operators}

The SM CPV in the weak sector leads to effective dimension-6
four-fermion operators at hadronic scales.  In principle, these also
lead to dimension-3 CPV mass terms, \(\bar\psi \gamma_5 \tau \psi\), for
the fermions, where \(\psi\) is the fermion field and \(\tau\) is a
flavor matrix\footpunct{We ignore possible CP-violating Majorana
  phases in the neutrino sector~\citep{PhysRevD.88.033002}}.  Axial
transformations can be used to remove the quark CPV masses, except when \(\tau\) is
the identity, in which case the anomaly transforms it to the
dimension-4 gluon-topological-charge operator (also called the
\(\Theta\)-term), \(G_{\mu\nu}\tilde G^{\mu\nu}\), where \(G\) is the
gluon field strength~\citep{PhysRevD.92.114026}. Phenomenological
estimates, using the limit on the neutron electric dipole moment, 
already constrain the total coefficient \(\bar\Theta\) of this
operator to be anomalously small, less than
\(3\times10^{-10}\)~\citep{POSPELOV2000177}.

In BSM theories, CPV operators start at dimension 6 at the weak
scale~\citep{POSPELOV2005119}, but two of them---the quark EDM (qEDM),
\(\bar\psi\tau\Sigma_{\mu\nu}\tilde F^{\mu\nu}\psi\), where \(F\) is
the electromagnetic field tensor, and the quark chromo-EDM (qCEDM),
\(\bar\psi\tau\Sigma_{\mu\nu}\tilde G^{\mu\nu}\psi\)---become
dimension 5 after electroweak symmetry breaking.  This means that
their natural suppression relative to the QCD scale is by \(v_{\rm
  EW}/M^2_{\rm BSM}\) rather than by \(1/M^2_{\rm BSM}\) as for the
remaining dimension-6 operators---the gluon chromo-EDM (also called
the Weinberg 3-gluon operator), \(G_{\mu\nu}G_{\lambda\nu}\tilde
G_{\mu\lambda}\), and various 4-fermion operators. In many BSM models,
however, the dimension-5 operators come with extra Yukawa suppression,
and their effect is comparable to the other dimension-6
operators. Thus, all these should be considered at the same level, and
their ME within the neutron state calculated.

%\subsection{Effective Field Theory}
%
%{\centering
%\leavevmode\colorbox{white}{\includegraphics[width=0.95\hsize]{schemetb2}}
%
%}
%
%{\hfill\tiny Pospelov and Ritz, {\it Ann. Phys.} {\bf 318} (2005) 119.}
%

\subsection{Form Factors}

Using Lorentz symmetry, the response of a neutron to the vector current can be written in terms of the 
Dirac \(F_1\), Pauli \(F_2\), anapole \(F_A\), and electric-dipole \(F_3\) form-factors as 
\begin{eqnarray*}
\langle N | V_\mu(q) | N \rangle & = &
   \overline {u}_N \left[ 
         \gamma_\mu\;F_1(q^2) + i \frac{[\gamma_\mu,\gamma_\nu]}2 q_\nu\; \frac{F_2(q^2)}{2 m_N} + {(2 i\,m_N \gamma_5 q_\mu - \gamma_\mu \gamma_5 q^2)\;\frac{F_A(q^2)}{m_N^2}} \right.\\[1\jot]
&&\qquad\qquad\qquad\qquad\qquad\qquad\qquad\qquad\qquad\left.
         {} + {\frac{[\gamma_\mu, \gamma_\nu]}2 q_\nu \gamma_5\;\frac{F_3(q^2)}{2 m_N}} \right] u_N\,, 
\end{eqnarray*}
in the Euclidean metric. Here \(V_\mu\) represents the
electromagnetic vector current, \(u_N\) represents the neutron spinor
normalized such that \(u_N \bar u_N = -i\pslash + m_N\), where \(p\)
is its momentum and \(m_N\) its mass, \(q\) the momentum inserted by
the vector current \(V_\mu\), and \(|N\rangle\) is the neutron state. The
Sachs form factors~\cite{Ernst:1960zza,BARNES1962166,Hand:1963zz} that
describe the charge and current densities in the Breit frame, are
related to these as \(G_E \equiv F_1 - (q^2/4M^2) F_2\) and \(G_M
\equiv F_1 + F_2\)\footpunct{In this frame, the form factor \(F_3\)
  contributes a spin-dependent charge-density, and \(F_A\), a
  current density.}. The electric charge is \(G_E(0) = F_1(0) = 0\) and
the anomalous magnetic dipole moment is \(G_M(0)/2 M_N = F_2(0) / 2
M_N\).  The anapole moment breaks the symmetry under simultaneous
parity-flip and time-reversal (PT), and so, will be zero in our
calculations.  The electric dipole moment is given by the CP-violation
form factor \(F_3\) at zero \(q^2\), \(d_E = F_3(0)/2 m_N\).\looseness-1

If parity is violated, an operator that creates an asymptotic neutron
with the standard parity transformation properties is
\(N_{\alpha_N}\equiv \epsilon_{abc}\left[(\bar d^a)^C \gamma_5 P
  u^b\right] \exp{\{i\alpha_N\gamma_5\}} d^c\), where \(a\), \(b\),
and \(c\) are color labels, the superscript \(C\) represents charge
conjugation, \(P\equiv(1+\gamma_4)/2\) is positive-energy projector
for zero-momentum quarks that improves the signal\footpunct{At nonzero
  momentum, this introduces a mixing with the spin-3/2 state, which
  being heavier than the neutron, is controlled like other excited
  states.}, and \(\alpha_N\) is a constant depending on the asymptotic
state that needs to be determined\footpunct{\(\alpha_N\) is also
  specific to the precise operator used: different operators with the
  same quark content and Lorentz properties can, in principle, need
  different \(\alpha_N\).}.  Under the standard choice of quark and
neutron parities, \(\alpha_N\) is real when PT is a good symmetry,
imaginary when CP is good, and zero if parity is unbroken.
\vspace*{-\baselineskip}
\section{Status of Lattice Calculations and Preliminary Results}
\vspace*{-0.5\baselineskip}
\begin{figure}[t]
  \begin{minipage}[t]{0.72\textwidth}
    \hrule width 0pt
    \begin{center}
      \setlength{\tabcolsep}{0pt}
	\begin{tabular}{cc}
		\colorbox{white}	
		{%
		\includegraphics[width=0.41\textwidth,viewport=8 2 247 172,clip]{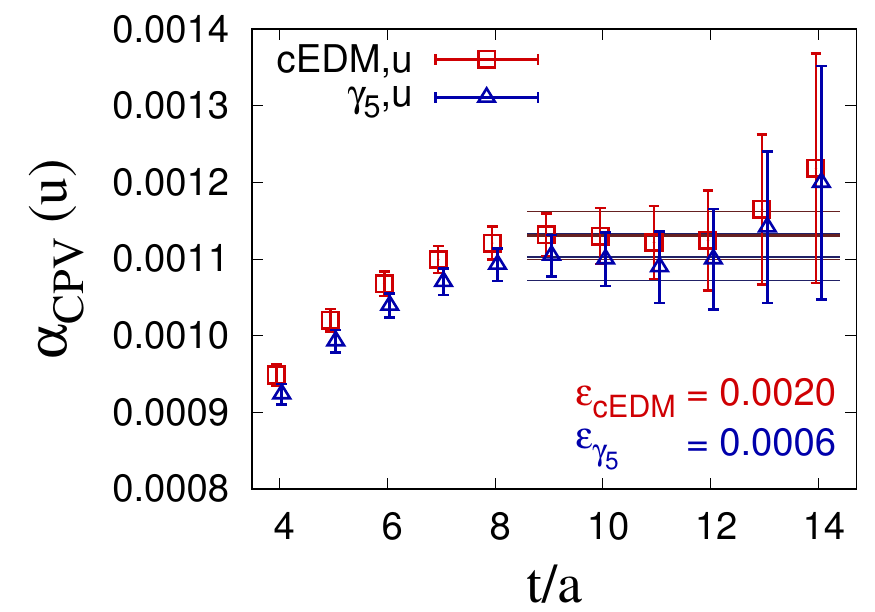}%
		}&%
	    \colorbox{white}{%
	    	\includegraphics[width=0.55\textwidth,viewport=3 3 286 201,clip]{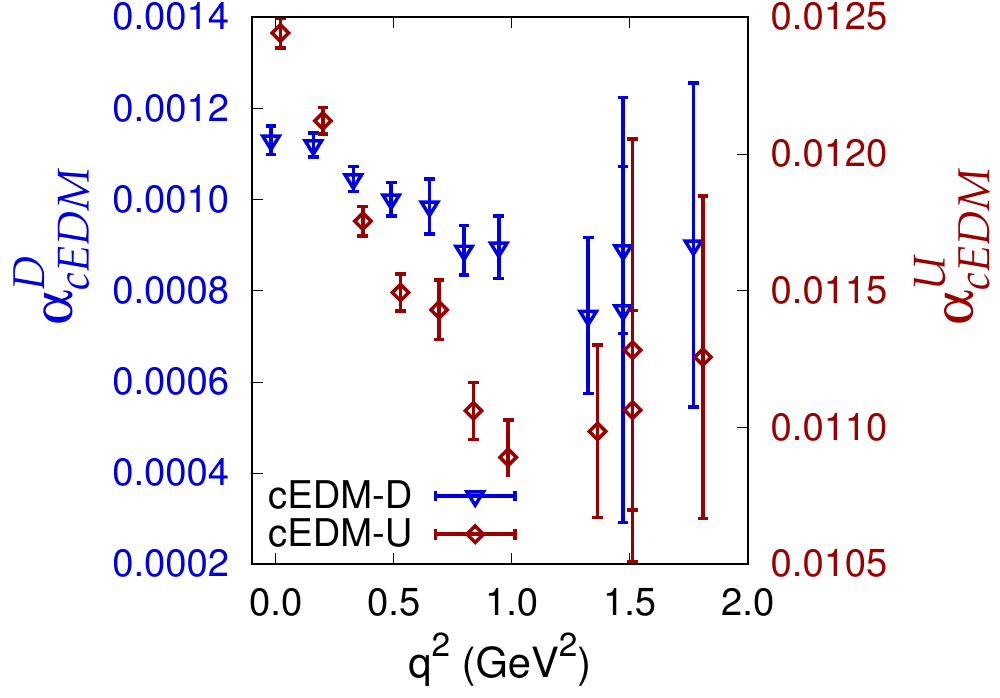}%
	    }\\[-1\jot]
	    (a)&(c)\\
		\colorbox{white}{%
		\includegraphics[width=0.41\textwidth,viewport=8 2 247 172,clip]{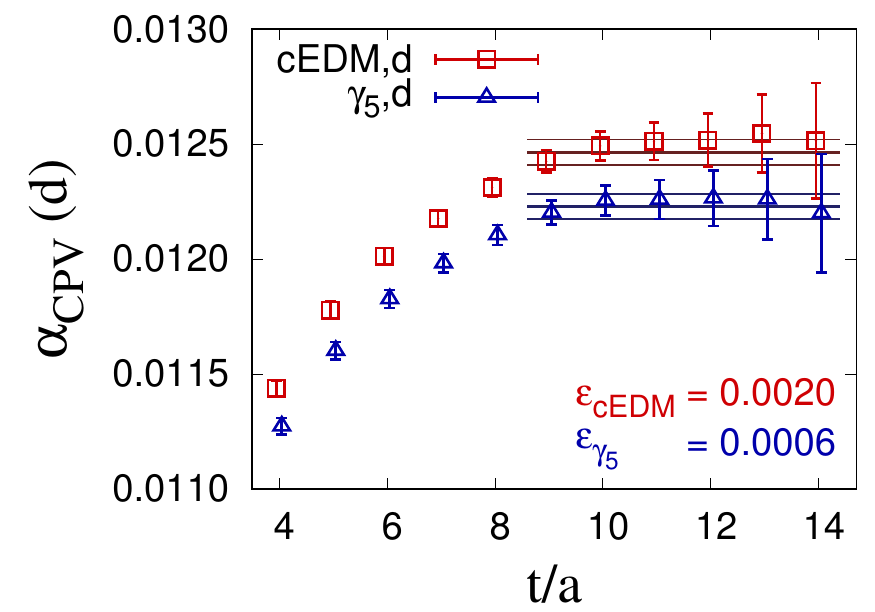}%
		}&%
	    \colorbox{white}{%
	    	\includegraphics[width=0.55\textwidth,viewport=3 3 286 201,clip]{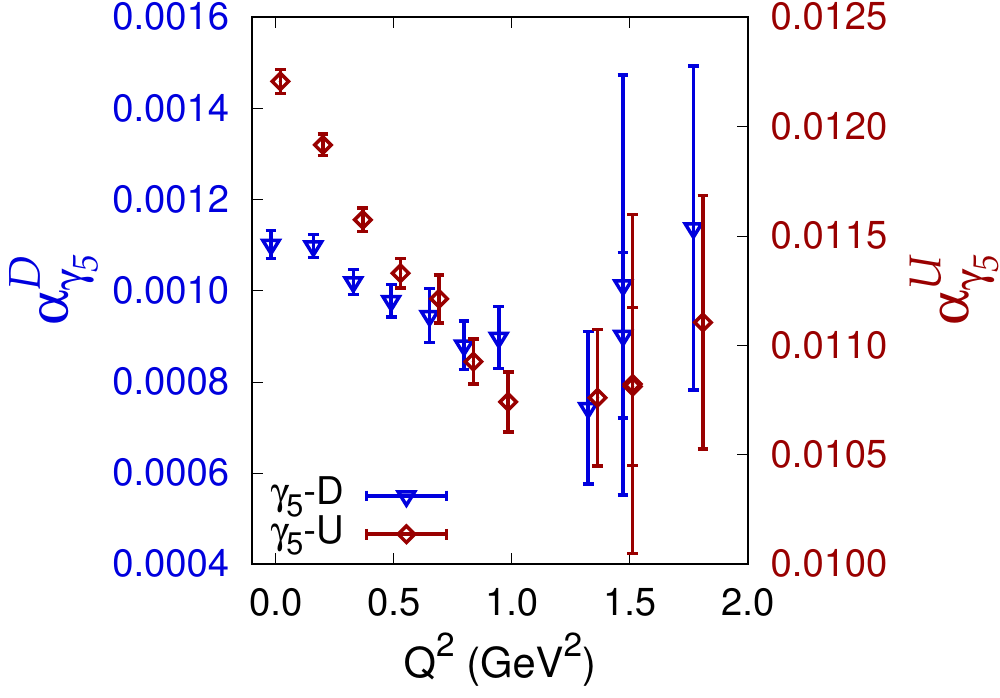}%
	    }\\[-1\jot]
        (b)&(d)
	\end{tabular}
      \end{center}
    \end{minipage}\hspace{0.01\textwidth}%
    \begin{minipage}[t]{0.25\textwidth}
	\caption{(a) The extraction of \(\alpha_N\) with the insertion of the chromo-EDM (labeled cEDM) or CPV mass (labeled \(\gamma_5\)) operators on \(u\) quarks. (b) Same as (a), but with the insertion on the \(d\) quark.  (c) \(\alpha_N\) due to chromo-EDM insertion as a function of the 3-momentum of the neutron; the insertions on the \(u\) and \(d\) quarks are shown separately. (d) Same as (c), but for the CPV mass term.\looseness-1}
	\label{fig:1}
      \end{minipage}
      \vspace*{-\baselineskip}
\end{figure}
We have recently completed an analysis of the proton and neutron EDMs
arising from the quark EDM~\citep{PhysRevD.98.091501}.  Here we report
on progress on nucleon EDMs arising from the quark chromo-EDM
operator.  This operator has power divergent mixing with lower
dimensional pseudoscalar quark mass term and the gluon topological charge
(if chiral symmetry is violated) that has to be controlled. We,
therefore, discuss these operators as well.\looseness-1

\begin{figure}[t]
  \begin{minipage}[t]{0.65\textwidth}
    \hrule width 0pt
    \begin{center}
      \setlength{\tabcolsep}{0pt}
    \begin{tabular}{cc}
      \colorbox{white}{%
      \includegraphics[width=0.45\hsize,viewport=16 6 229 184,clip]{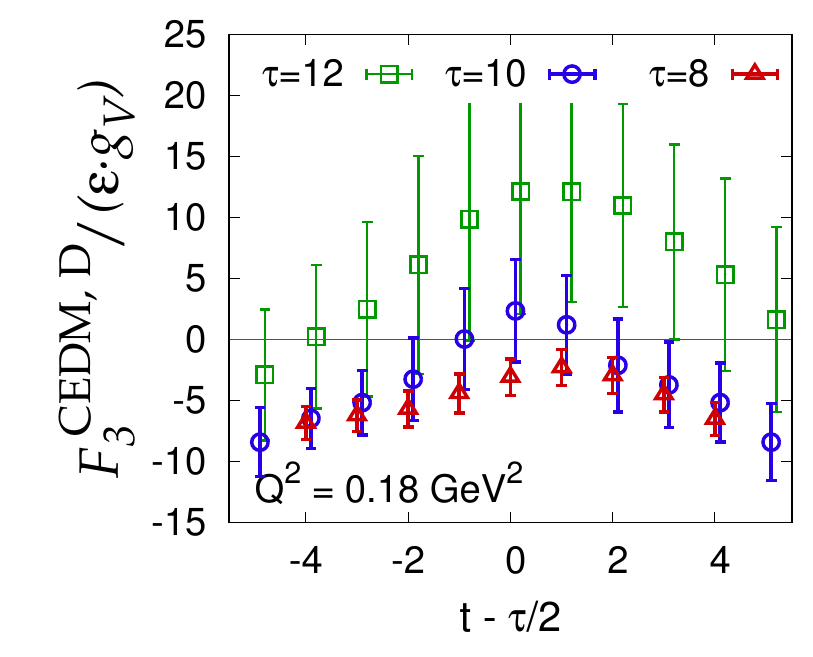}%
      }&%
         \colorbox{white}{%
         \includegraphics[width=0.45\hsize,viewport=16 6 229 184,clip]{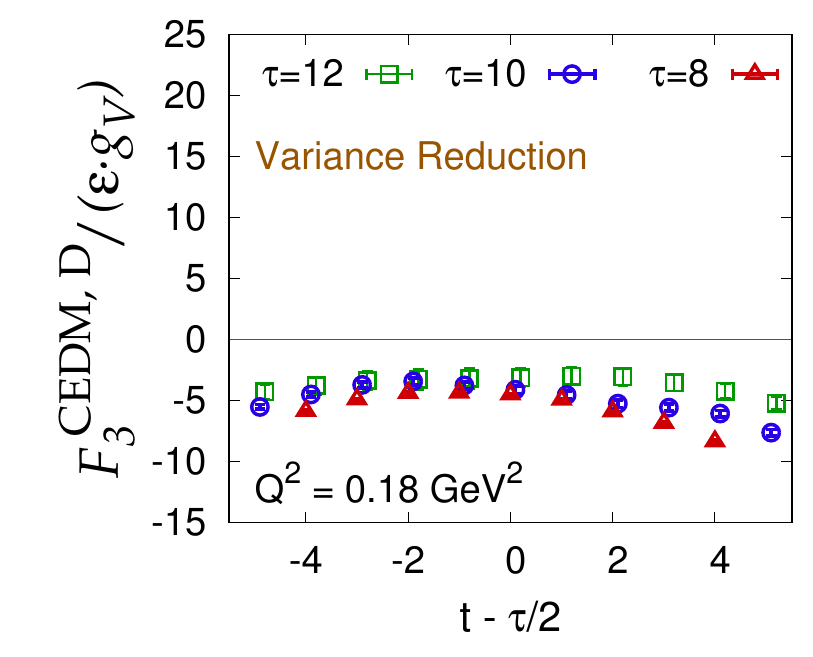}%
         }\\[-1\jot]
      (a)&(b)\\
      \colorbox{white}{%
      \includegraphics[width=0.45\hsize,viewport=16 6 229 184,clip]{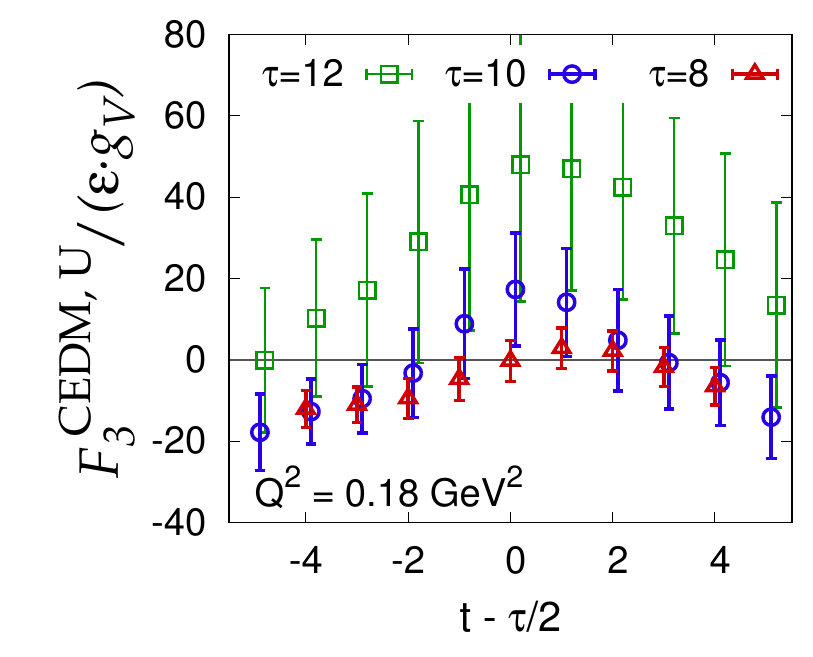}%
      }&%
         \colorbox{white}{%
         \includegraphics[width=0.45\hsize,viewport=16 6 229 184,clip]{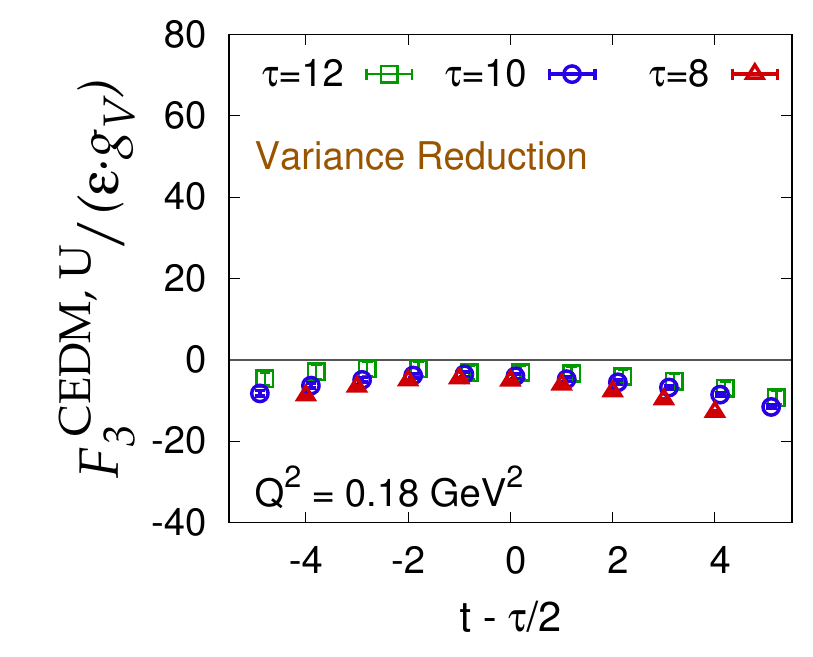}%
         }\\[-1\jot]
      (c)&(d)
    \end{tabular}
  \end{center}
\end{minipage}%
\begin{minipage}[t]{0.35\textwidth}
  \caption{\(F_3/(\varepsilon \cdot g_V)\) for the neutron due to chromo-EDM insertion on the \(d\) ((a) and (b)) and \(u\) ((c) and (d)) quarks before ((a) and (c)) and after ((b) and (d)) variance reduction. Here \(\varepsilon\) is the coefficient of the CP-violating term, and the baryonic vector charge \(g_V\) is unity in the continuum limit and cancels the lattice renormalization of the vector current.}
  \label{fig:3}
\end{minipage}
\vspace*{-\baselineskip}
\end{figure}
%\subsection{Technique}
The quark chromo-EDM operator is a quark bilinear, so its ME can be calculated using 
the Schwinger source method as described in previous
proceedings~\citep{Bhattacharya:2016rrc,refId0}.  
All the calculations presented here were done with Clover fermions on
a MILC generated \(L^3 \times T = 24^3 \times 64\) HISQ
ensemble~\citep{PhysRevD.75.054502,PhysRevD.87.054505} with lattice
spacing \(a = 0.1207(11)~\mbox{fm}\) and pion mass \(M_\pi^{\rm sea} =
305.3(4)~\mbox{MeV}\). The parameters of the Wilson-clover action used
are \(\kappa \approx 0.1272103\) corresponding to \(M_\pi^{\rm val} =
310.2(2.8)~\mbox{MeV}\) and \(c_{\rm SW} = 1.05094\) determined using
tree-level boosted perturbation theory with \(u_P^{HYP} =
0.9358574(29)\).  Statistical precision is increased using the
truncated solver method~\citep{BALI20101570,PhysRevD.88.094503} with
128 low-precision and 4 high-precision measurements on each of the
1012 configuration.  The quark-disconnected diagrams were ignored in
all of the reported calculations.\looseness-1

To determine \(\alpha_N\), we use  \(\Im\Tr \gamma_5
\langle N_0 \bar N_0\rangle/\Re\Tr \langle N_0 \bar N_0\rangle\),
where \(\langle N_0 \bar N_0\rangle \) is the propagator with
\(\alpha_N=0\). Asymptotically, this gives \(-2 \tan\alpha_N\). 
Figures~\ref{fig:1}(a) and (b) show \(\alpha_N\) extracted 
from the neutron propagator for the CP
violation parameter \(\varepsilon\) in the small (linear) 
regime~\cite{Bhattacharya:2016rrc}.
% with the choice of \(\varepsilon_{\textrm{cEDM}}=0.002\) and \(\varepsilon_{\gamma_5}=0.0006\).
% In reference \cite{Bhattacharya:2016rrc}, It is shown that these values of \(\varepsilon\) are in the linear region.
Figures~\ref{fig:1}(c) and (d) show that there is a strong dependence
of the extracted \(\alpha_N\) on the momentum of the neutron, probably
due to lattice spacing artifacts.\looseness-1

%\section{Three-point Functions}
%
%\subsection{Projection}

\begin{figure}[t]
\begin{center}
\colorbox{white}{%
\includegraphics[width=0.4\textwidth]{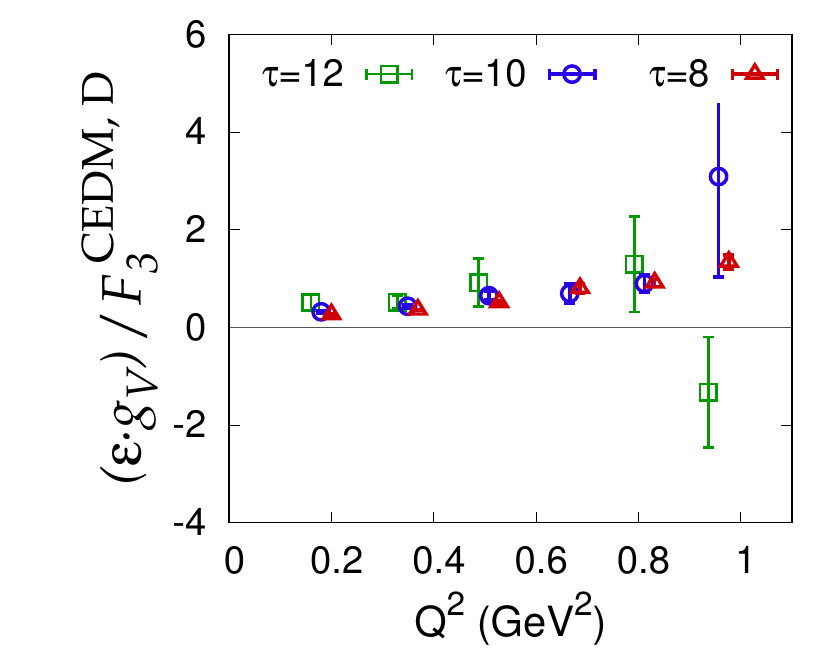}%
\hspace{0.05\textwidth}
\includegraphics[width=0.4\textwidth]{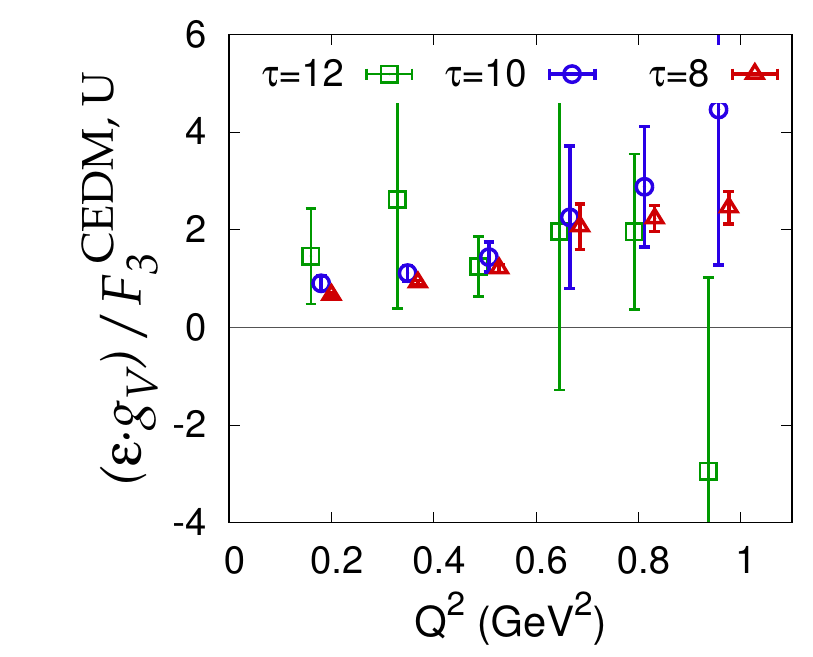}%
}
% \colorbox{white}{%
% \includegraphics[width=0.3\textwidth]{data/F3D_inv_cEDM_a12m310}%
% \hspace{0.05\textwidth}
% \includegraphics[width=0.3\textwidth]{data/F3U_inv_cEDM_a12m310}%
% \hspace{0.05\textwidth}{Proton }
% }\\
\end{center}
\vspace*{-0.8cm}
\caption{\((\varepsilon \cdot g_V)/F_3\) versus $Q^2$ for the neutron with
  chromo-EDM insertion on the \(d\) (left) and \(u\) (right) quarks at
  various source-sink separations. }
\label{fig:4}
\vspace*{-\baselineskip}
\end{figure}

The three point function, from which the ME of $V_\mu$ are extracted, calculated is
\begin{eqnarray*}
\langle \Omega | {N_0(\vec0,0) V_\mu(\vec q,t) \overline{N_0}(\vec p,T)} | \Omega \rangle &=&
\sum_{n,n'} e^{-i\alpha_{n}\gamma_5} u_n e^{-m_n t}\; {\langle n | V_\mu(q) | n' \rangle}\; e^{-E_{n'} (T-t)}{\overline u}_{n'} e^{-i\alpha_{n'}^*\gamma_5}
\end{eqnarray*}%
with projection onto only one spinor component using  
\( {\cal P}\equiv \frac12 (1+\gamma_4) (1+i\gamma_5\gamma_3) \). The 
ME is isolated using the combination
\[
R^\mu \equiv \frac{C_{\rm 3pt}^\mu (q;\tau,t)}{C_{\rm 2pt}(\tau;0)}
  \sqrt{\frac{C_{\rm 2pt}(t;0)C_{\rm 2pt}(\tau; 0) C_{\rm 2pt}(\tau-t;-q)}
             {C_{\rm 2pt}(t;-q)C_{\rm 2pt}(\tau;-q) C_{\rm 2pt}(\tau-t;0)}}\,
\]
where \(C_{\rm 3pt}^\mu\) are the projected 3pt functions and \(C_{\rm 2pt}\) are the real parts of projected 2pt functions.
\iffalse
One then has
\begin{eqnarray*}
F_3 = - \frac{2 m\sqrt{2 E}}{\cos\alpha[q_3\sqrt{E+m}(q_1^2+q_2^2)(E+m\cos 2\alpha)]} \times {}\span\span\\
\span\span {} \times
       \left[\cos\alpha (E+m)^2 (q_1 \Re V_1 + q_2 \Re V_2) + q_3 \sin\alpha [(E+m) (q_2 \Re V_1 - q_1 \Re V_2)
                                           + (q_1^2+q_2^2)\Re V_4]\right]\,,
\end{eqnarray*}
where \(V_i\) are the isolated matrix elements.  
\fi 
%
The eight quantities---$\Re R^\mu$ and $\Im R^\mu$---provide an overdetermined set from which 
the three form factors can be extracted\footpunct{Note, we also consider \(R\) with
  momentum components permuted and reflected according to the
  symmetries of the theory, but do not display them explicitly in the
  narrative.}.  In fact, these break up into two sets of four
quantities: the components \(V_R\equiv(\Re \vec R, i\Im R^4)\) give 
 \(G_2 \equiv G_E + \tan\alpha_N (Q^2/4m_N^2) F_3\), whereas
the other four components \(V_I \equiv (\Im \vec R, -i \Re R^4)\)
give  \(G_1 \equiv G_M\) and \(G_3 \equiv F_3 + \tan\alpha_N
F_2\)\footpunct{One can also account for possible current
  nonconservation by including two additional form factors, a
  combination of which appears in each of the sets.  The effects of
  including these neglected terms were found to be
  small. We also ignore purely lattice form factors (i.e.,
  coefficients of hypercubic covariant, but not Lorentz covariant,
  tensors) including those arising from violation of the relativistic
  dispersion relation.}.  The overdetermined set of equations for the transition ME between a neutron at 
rest and momentum \((-\vec q,E_N)\) is 
\[
 E\equiv \begin{pmatrix}
    X_1&0&X_3\\
    0&Y_1&0
  \end{pmatrix}
  \begin{pmatrix}
    G_1\\
    G_2\\
    G_3
  \end{pmatrix} - \begin{pmatrix} V_R\\V_I\end{pmatrix}=0\,,
\]
where \(X_1 \equiv m (-cq_2, cq_1, s(E_N-m_N), -isq_3)^T\), \(Y_1
\equiv m_N c (q_1, q_2, q_3, -i(E_N+m_N))^T\), \(X_2\equiv
-(q_3/2m_N)Y_1\), \(c\equiv \cos^2\alpha_N\) and \(s\equiv
\sin\alpha_N\cos\alpha_N\). We solve this set by the method of least squares, i.e., we solve the linear
equations obtained by differentiating \(E^T W E\) with respect
to \(G_i\) for a positive weight matrix \(W\),  which we choose, for
simplicity, to be \(\diag(\sigma_R^2,\sigma_I^2)^{-1}\), where
\(\sigma_R\) and \(\sigma_I\) are the errors on \(V_R\) and \(V_I\)
respectively.\looseness-1

%\subsection{\relax\texorpdfstring{$F_3$}{F\textsubscriptthree} Form factor from CEDM}

Figure~\ref{fig:3} (a) and (c) show that the signal in the \(F_3^u\)
and \(F_3^d\) is, a priori, poor. To improve the signal, we propose
the following variance reduction method: use quantities \(z_i\) that
are correlated with \(F_3\) and have zero expectation value, and
construct the lower variance estimator \(F_3 - \sigma^2_{Fi}
\Sigma^{-2}_{ij} z_j \), where \(\Sigma^{-2}\) is the inverse
variance-covariance matrix of \(z_i\) and \(\sigma^2_{Fi}\) is the
covariance of \(F_3\) with \(z_i\).  Since we know that for
\(\varepsilon =0\) (CP-symmetric case), \(F_3\) is zero even at finite
$a$ and for our mixed action calculation, and remains highly correlated with
\(F_3\) for small \(\varepsilon\), we expect using 
\(F_3(\varepsilon=0)\) as a \(z_i\) in the above expression will reduce
the variance.  \looseness-1
Figures~\ref{fig:3} (b) and (d) show that this variance reduction method improves the signal substantially.

Figure~\ref{fig:4} shows that \(1/F_3\) behaves linearly with \(Q^2\)
as would be expected from pole-dominance. Also, the dependence on the
source-sink separation $\tau$ is small, indicating excited state
contributions are manageable. The next important step, once the signal
is established, is to subtract power divergences in the
renormalization due to operator mixing to ensure a finite result in
the continuum limit.

\begin{figure}[t]
\begin{center}
\colorbox{white}{%
\includegraphics[width=0.4\hsize]{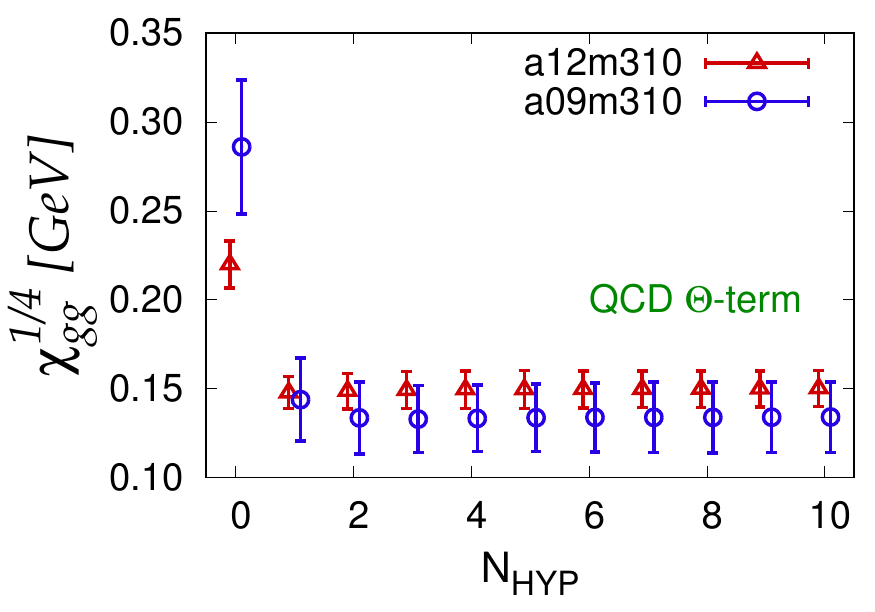}%
}
\hspace{0.1\hsize}%
\colorbox{white}{%
\includegraphics[width=0.4\hsize]{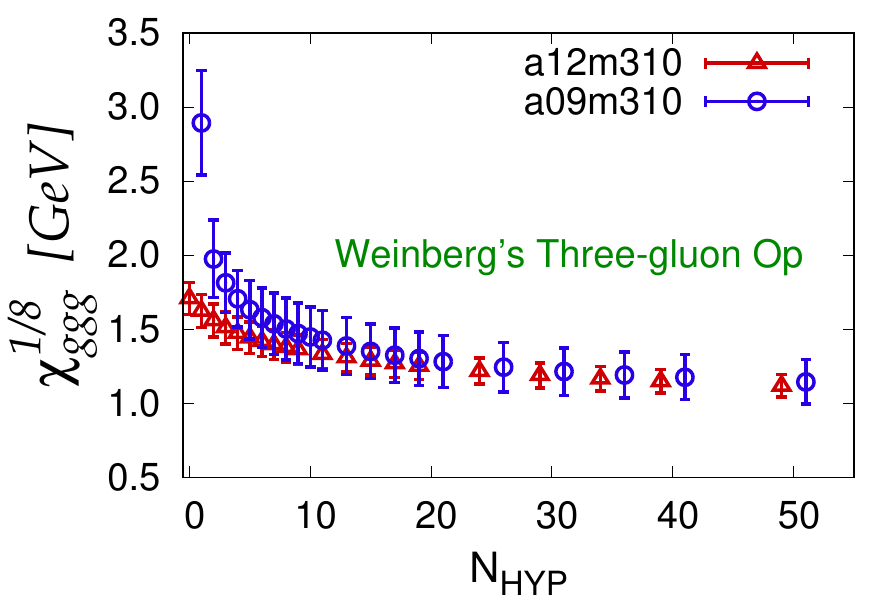}%
}
\end{center}
\vspace*{-0.8cm}
\caption{The evolution of the susceptibilities of the topological
  charge (left) and the Weinberg operator (right) as a function of the HYP 
  smearing steps.}
\label{fig:6}
      \vspace*{-0.5\baselineskip}
\end{figure}

\begin{figure}[t]
  \begin{minipage}{0.4\hsize}
\begin{center}
\colorbox{white}{%
\includegraphics[width=0.9\hsize]{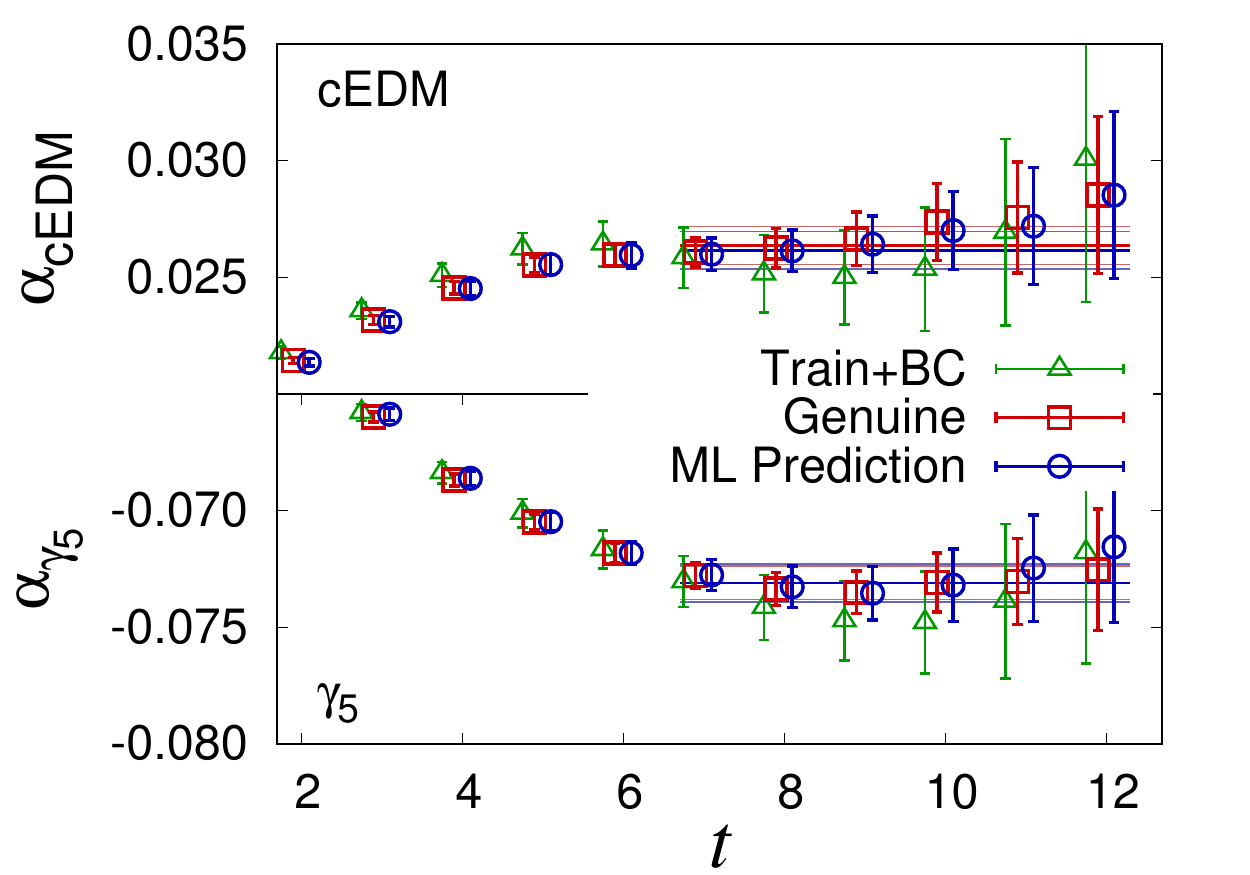}%
}
\end{center}
\end{minipage}
\begin{minipage}{0.58\hsize}
  \caption{Comparison of the predicted and calculated imaginary part
    of the propagator with the insertion of the chromo-EDM and CPV
    mass term from the real part of the propagator with no
    insertion. Note that the final errors on the prediction are
    comparable to the actual errors, substantially reduced from the
    errors on the train set. These results are obtained using the
    choice of \(\varepsilon_{\textrm{cEDM}}=0.004\) and
    \(\varepsilon_{\gamma_5}=-0.004\).}
\label{fig:5}
\end{minipage}
\vspace*{-\baselineskip}
\end{figure}

\vspace*{-\baselineskip}
\section{Ongoing Work}
\vspace*{-0.5\baselineskip}

We are extending the calculation of the chromo-EDM by constructing
various combinations of the matrix elements that each provide an estimate
of \(F_3\),  and by adding the contribution of the disconnected diagrams.
To address divergent mixing under renormalization, the RI-sMOM scheme defined
in~\citep{PhysRevD.92.114026} is complicated, so we are evaluating
gradient-flow regularization~\citep{Luescher2010}. We will first
implement this scheme for the simpler case of the two gluonic
operators. In Figure~\ref{fig:6}, we show the evolution of the
susceptility of the scale-dependent Weinberg operator with the number
of HYP smearings and compare it to the scale-independent topological
susceptibility.  We are investigating whether the expected
scale-dependence persists under gradient-flow smearing. The
motivation is that under gradient flow renormalization, the Weinberg
operator needs no divergent subtraction.  The next step will be to
extend this to the fermion sector. To demonstrate efficacy, we will 
first compare results for isovector charges renormalized using the RI-sMOM
scheme and gradient flow. \looseness-1

Finally, we are developing machine learning algorithms to reduce the
computational effort. Building on the similarity to reweighting
ensembles or unraveling quantum trajectories, the method proceeds by
finding a combination of easily calculated and statistically precise
quantities that have a high correlation with more compute-intensive
quantities of interest, and then making the estimates rigorous by
implementing standard bias reduction techniques. Initial tests of
these ideas show promise as illustrated in Figure~\ref{fig:5}.

We acknowledge computer resources provided by LANL (IC), NERSC (US DOE contract  DE-AC02-05CH11231), OLCF (US DOE contract DE-AC05-00OR\-22725) and JLAB (USQCD); the CHROMA software suite~\cite{EDWARDS2005832}; and support from US DOE contract DE-AC52-06NA25396 and LANL LDRD grant 20190041DR.

\vspace*{-0.5\baselineskip}

\bibliographystyle{naturemodified}

\bibliography{lat18}

%\begin{thebibliography}{99}
%\bibitem{...}
%....
%
%\end{thebibliography}

\end{document}